\documentclass[twocolumn,eqsecnum,prb,aps,endfloats]{revtex4}
\usepackage{graphicx}
\newcommand{\e}{{\rm e}}
\newcommand{\ep}{\varepsilon}
\newcommand{\ld}{\lambda}

\newcommand{\nn}{\nonumber}

\newcommand{\be}{\begin{equation}}
\newcommand{\ee}{\end{equation}}
\newcommand{\ba}{\begin{eqnarray}}
\newcommand{\ea}{\end{eqnarray}}
\def\ft#1#2{{#1\over #2 }}
\begin{document}
\title{The Structure of Positive Decompositions of Exponential Operators}

\author{Siu A. Chin}

\affiliation{Department of Physics, Texas A\&M University,
College Station, TX 77843, USA}

\begin{abstract}
The solution of many physical evolution equations can be expressed
as an exponential of two or more operators acting on initial 
data. Accurate solutions can be systematically derived by 
decomposing the exponential in a product form. For time-reversible 
equations, such as the Hamilton or the Schr\"odinger equation, it is 
immaterial whether or not the decomposition coefficients are positive. 
In fact, most symplectic algorithms for solving classical dynamics 
contain some negative coefficients. For time-irreversible systems, 
such as the Fokker-Planck equation or the quantum statistical propagator,
only positive-coefficient decompositions, which respect the 
time-irreversibility of the diffusion kernel, can yield practical 
algorithms. These positive time steps only, forward decompositions, 
are a highly effective class of factorization algorithms. This work 
introduce a framework for understanding the structure of these 
algorithms. By a suitable representation of the factorization 
coefficients, we show that specific error terms and order conditions can
be solved {\it analytically}. Using this framework, we can go beyond 
the Sheng-Suzuki theorem and derive a lower bound for
the error coefficient $e_{VTV}$. By generalizing the
framework perturbatively, we can further prove that it is not possible 
to have a sixth order forward algorithm by including only the 
commutator $[VTV]\equiv[V,[T,V]]$. The pattern of these higher
order forward algorithms is that in going from
the (2n)$^{\rm th}$ to the (2n+2)$^{\rm th}$ order, one must 
include a new commutator $[VT^{2n-1}V]$ in the 
decomposition process. 

\end{abstract}
\maketitle

\section {Introduction}
Many physical evolution equations, from classical 
mechanics\cite{yoshi,hairer,mcl02,chinchen03}, 
electrodynamics\cite{hirono},
statistical mechanics\cite{ti,chincor} to quantum 
mechanics\cite{feit,chinchen01,chinchen02}, 
all have the form
\be
{{\partial w}\over{\partial t}}=(T+V)w,
\label{gen}
\ee
where $T$ and $V$ are non-commuting operators. Such an equation
can be solved iteratively via
\be
w(t+\epsilon)={\rm e}^{\epsilon(T+V)} w(t),
\label{two}
\ee
provided that one has a suitable approximation for the short
time evolution operator ${\rm e}^{\epsilon(T+V)}$. Usually,
${\rm e}^{\epsilon T}$ and ${\rm e}^{\epsilon V}$ can be solved
exactly. By factorizing  ${\rm e}^{\epsilon(T+V)}$
to higher order in the form
\be
{\rm e}^{\epsilon (T+V )}=\prod_{i=1}^N
{\rm e}^{t_i\epsilon T}{\rm e}^{v_i\epsilon V},
\label{prod}
\ee
one can solve (\ref{gen}) accurately with excellent conservation 
properties. Classically, each factorization \ref{prod}) produces a 
{\it symplectic integrator} which exactly conserve all
Poincar\'e invariants. A vast literature\cite{yoshi,hairer,mcl02} 
exists on producing symplectic integrators of the form (\ref{prod}). 
Once a factorization scheme is derived, it can be implemented
specifically to solve any particular evolution equation of the 
form (\ref{gen}).   

However, as one examines these factorization schemes more closely, 
one is immediately struck by the fact that beyond second order, all
such scheme contain some negative coefficients\cite{yoshi,hairer,mcl02}
$t_i$ and $v_i$. Since the fundamental 
diffusion kernel cannot be simulated or integrated backward in 
time, none of these higher order schemes can be applied to 
{\it time-irreversible} systems. This lack of positive-coefficient 
decompositions beyond second order was first noted and proved 
by Sheng\cite{sheng}. Sheng showed that equations for determining the 
third order coefficients in (\ref{prod}) are incompatible if the
coefficients $\{t_i,v_i\}$ are assumed to be positive. This is a
valuable demonstration, but it shed no light on the cause of this
incompatibility nor offered clues on how to overcome this
deficiency. Suzuki\cite{suzukinogo} later proved that the
incompatibility can be viewed more geometrically. His proof tracked
the coefficients of the operator $TTV$ and $TVV$ in the product
expansion of (\ref{prod}). If the expansion were correct to third
order, then the coefficients for both operators must be $1/3!$. The
coefficient condition for one corresponds to a hyperplane and the
other, a hypersphere. Suzuki then went on to show that for the same
set of positive coefficients, the hyperplane cannot intersect the
hypersphere and therefore no real solution is possible.

The product form (\ref{prod}) has the general expansion
\ba
\prod_{i=1}^N&&
\e^{t_i\ep T}\e^{v_i\ep V}
=\exp\biggl( e_T \ep T+e_V \ep V+e_{TV}\ep^2[T,V]\nonumber\\
&&+\,e_{TTV}\ep^3[T,[T,V]]+e_{VTV}\ep^3[V,[T,V]]+\cdots\biggr)\nn\\
&&=\e^{\ep H_A(\ep)}, 
\label{prodform} 
\ea 
where the last equality
defines the approximate Hamiltonian of the product decomposition.
The goal of factorization is to keep $e_T=e_V=1$ and forces all
other error coefficients such as $e_{TV}$, $e_{TTV}$, $e_{VTV}$,
{\it etc.}, to zero. By tracing the incompatibility condition to
error coefficients of specific operators, one can identify which
error term cannot be made to vanish. The operator $TTV$ can only
occur in $[T,[T,V]]$ and $TVV$ only in $[V,[T,V]]$. Thus the
incompatibility condition is equivalent to the fact that for
positive coefficients $\{t_i,v_i\}$, $e_{TTV}$ and $e_{VTV}$
cannot both be reduced to zero. To circumvent this, it is suffice
to force one error coefficient to zero and keep the other
commutator in the factorization process. Since in quantum
mechanics $[V,[T,V]]$ corresponds to a local function, just like
the potential, Suzuki\cite{suz95} suggested that one should
factorize $\e^{\ep (T+V)}$ in terms of $T$, $V$ {\it and}
$[V,[T,V]]$. Following up on this suggestion, Suzuki\cite{suzfour}
and Chin\cite{chin} have derived fourth order factorization
algorithms with only positive coefficients. Chin\cite{chin} also
shown that classically, $[V,[T,V]]$ give rises to a 
force gradient exactly as first suggested by Ruth\cite{ruth83}. 
Chin and collaborators have since abundantly demonstrated the 
efficiency of these forward time step algorithms in solving both
time-irreversible\cite{fchinl,fchinm,auer,ochin} and
time-reversible\cite{chin,chinchen01,chinchen02,chinchen03}
dynamical problems. Jang {\it et al.}\cite{jang} have used
these forward factorization schemes in doing 
quantum statistical calculations and Omelyan 
{\it et al.}\cite{ome02,ome03} have
produced an extensive collection of higher order algorithms 
based on this class of fourth order forward algorithms.

An important question therefore arises: with the inclusion of the
operator $[V,[T,V]]$, can one produce forward algorithms
of sixth or higher order? The answer provided by this work is
``no". For a sixth order decomposition with positive coefficients,
the commutator $[V,[T,[T,[T,V]]]]$ cannot be made to vanish and
must be included. In order to prove this result we have developed 
a formalism to analyze the structure of these forward factorization
schemes. By use of a suitable representation of the
factorization coefficients, we show that linear
order conditions and quadratic error terms can both be solved 
{\it analytically}. The resulting error term then makes it obvious
that it cannot vanish if the factorization coefficients are purely 
positive. By use of this formalism we can go beyond
the Sheng-Suzuki theorem and derive a lower bound for the
magnitude of the error coefficient $e_{VTV}$. By generalizing the 
method to sixth order, we further prove the main result as 
stated above. This analytical method of 
solving the order conditions will allows us to analyze and classify 
factorization algorithms in general.

In the next section we introduce our notations and illustrate our 
method of solving the order condition analytically by giving a 
constructive proof of the Sheng-Suzuki theorem.
In Section III, we discuss the conditions necessary for a six order 
forward algorithm. In Section IV we introduce a perturbative
approach to study the sixth order case and
show that it is not possible to have a forward
sixth order algorithm by including only the commutator
$[V,[T,V]]$. In Section V we discuss the pattern of higher 
order forward algorithms. In Section VI, we summarize
our conclusions and suggest directions for future research.
The Appendix contains details of how to reduce a general
quadratic error coefficient to a multi-diagonal form.

\section{A constructive proof of the Sheng-Suzuki theorem}

In Suzuki's proof\cite{suzukinogo}, without explicitly computing 
$e_{TTV}$ and $e_{VTV}$, he showed that both cannot be zero.
Here, we show that by enforcing
$e_{TV}=0$ and $e_{TTV}=0$, we can compute a lower bound
for $e_{VTV}$ analytically and show that it cannot vanish for a set 
of positive $\{t_i\}$. This determination of a lower
bound for $e_{VTV}$ goes beyond the Sheng-Suzuki theorem in
providing a more detailed understanding of all fourth order
forward algroithms.  

The first step of our approach is to compute the error coefficients
$e_{TV}$, $e_{TTV}$, $e_{VTV}$, {\it etc.}, in terms of
the factorization coefficients $\{t_i,v_i\}$. This can be done as
follow. The left hand side of (\ref{prodform}) can be expanded as
\ba 
{\rm e}^{\ep t_1 T}
{\rm e}^{\ep v_1 V}
&
\cdots
&
{\rm e}^{\ep t_N T}
{\rm e}^{\ep v_N V}
=1+\ep \left (
\sum_{i=1}^N t_i\right ) T\nn\\
&&
+\ep \left( \sum_{i=1}^N
v_i\right) V +\cdots.
\label{rhoalh} 
\ea
Fixing $e_T=e_V=1$, the right hand side of (\ref{prodform}) 
can likewise be expanded
\ba
\e^{\ep H_A(\ep)}=&&1+\ep(T+V)+{1\over 2}\ep^2(T+V)^2
+\ep^2 e_{TV}[T,V]\nn\\
&&+\ep^3 e_{VTV}[V,[T,V]]+\ep^3e_{TTV}[T,[T,V]]\nn\\
&&+{1\over 2}\ep^3e_{TV}
\left \{(T+V)[T,V]+[T,V](T+V)\right \}\nonumber\\
&&+{1\over {3!}}\ep^3(T+V)^3+\cdots.
\label{rhoarh}
\ea
Matching the first order terms in $\ep$ gives the primary constraints
\begin{equation}
\sum_{i=1}^N t_i=1\quad{\rm and}\quad \sum_{i=1}^N v_i=1.
\label{tvcon}
\end{equation}
To determine the other error coefficients, we
focus on a particular operator in (\ref{rhoarh}) whose coefficient
contains $e_{TV}$, $e_{TTV}$ or $e_{VTV}$ and match that
operator's coefficients in the expansion of (\ref{rhoalh}). For
example, in the $\ep^2$ terms of (\ref{rhoarh}), the coefficient
of the operator $TV$ is $({1\over 2}+e_{TV})$. Equating this to
the coefficients of $TV$ from (\ref{rhoalh}) gives 
\be {1\over2}+e_{TV}=\sum_{i=1}^N s_i v_i, 
\label{etv} 
\ee 
where we have
introduced the variable 
\be s_i=\sum_{j=1}^i t_j. 
\label{si} 
\ee
Alternatively, the same coefficient
can also be expressed as 
\be {1\over 2}+e_{VT}=\sum_{i=1}^N t_i
u_i. 
\label{evt} 
\ee where 
\be u_i=\sum_{j=i}^N v_j. 
\label{ui}
\ee
It turns out that $s_i$ and $u_i$ are our fundamental variables,
the coefficients $t_i$ and $v_i$ are $backward$ and $forward$
finite differences of $s_i$ and $u_i$,
\ba
t_i&=&s_i-s_{i-1}\equiv\nabla\! s_i\nonumber\\
v_i&=&u_i-u_{i+1}\equiv-\nabla\!u_i
\label{tvdef}
\ea
The results (\ref{etv}) and (\ref{evt}) are equivalent by virtue of
the ``partial summation" identity
\be
\sum_{i=1}^N \nabla\! s_i u_i=-\sum_{i=1}^N s_i \Delta u_i.
\ee
(Note that $s_0=0$ and $u_{N+1}=0$.) In the following, we
will use the backward finite difference operator extensively,
\be
\nabla\!s_i^n=s_i^n-s_{i-1}^n,
\label{nab}
\ee
with property
$$
\sum_{i=1}^{N}\nabla\!s_i^n=s_N^n=1.
$$
Matching the coefficients of
operators $TTV$ and $TVV$ gives 
\ba &&{1\over 3!}+{1\over
2}e_{TV}+e_{TTV}={1\over 2}\sum_{i=1}^N s_i^2 v_i ={1\over
2}\sum_{i=1}^N\nabla\!s_i^2 u_i,\qquad\quad
\label{ettv}\\
&&{1\over 3!}+{1\over 2}e_{TV}-e_{TVT}
={1\over 2}\sum_{i=1}^N\nabla\!s_i u_i^2.
\label{evtv}
\ea
The error coefficient $e_{VTV}$ can be tracked directly by the 
operator $VTV$. The coefficient for the operator $VTV$ is quadratic 
in $v_i$ but not diagonal.  This is more difficult to deal with than 
$TVV$'s coefficient. Nevertheless,
we show in the Appendix that, $VTV$'s coefficient can be 
diagonalize by a systematic procedure to yield the same constraint
equation as (\ref{evtv}). 

In order to have a fourth order algorithm, aside from the 
primary constraints (\ref{tvcon}), one must require $e_{TV}=0$, 
$e_{TTV}=0$, and $e_{VTV}=0$. For a symmetric product form 
such that $t_1=0$ and $v_i=v_{N-i+1}$, $t_{i+1}=t_{N-i+1}$, 
or $v_N=0$ and $v_i=v_{N-i}$, $t_{i}=t_{N-i+1}$,
one has
\be
\e^{-\ep H_A(-\ep)}\e^{\ep H_A(\ep)}=1.
\ee
This implies that $H_A(\ep)$ must be a even function of $\ep$,
and $e_{TV}=0$ is automatic. The vanishing of  
all odd order errors in $H_A(\ep)$ implies that we must have 
\be
{1\over{(2n-1)!}}\sum_{i=1}^N\nabla\!s_i^{2n-1} u_i
={1\over{(2n)!}},
\label{eodd}
\ee
ensuring that $T^{2n-1}V$ has the correct expansion coefficient.
It is cumbersome to deal with symmetric coefficients directly,
it is much easier to use the general form (\ref{prod}) and to invoke
(\ref{eodd}) when symmetric factorization is assumed.

The next step in our strategy is compute a lower bound for
the magnitude of $e_{VTV}$, after satisfying 
constraints $e_{TV}=0$ and $e_{TTV}=0$. 
We view latter two constraints
\ba 
\sum_{i=1}^N\nabla\!s_i u_i&=&{1\over{2}},
\label{cone}\\
\sum_{i=1}^N\nabla\!s_i^2 u_i&=&{1\over{3}}, 
\label{ctwo} 
\ea
as constraints on $\{u_i\}$ for given a set of $\{t_i\}$ coefficients.  
For positive $\{t_i\}$, the RHS of (\ref{evtv}) is a
positive-definite quadratic form in $u_i$. Its lower bound
can be determined by the method of constrained minimization using 
Lagrange multipliers. Minimizing 
\ba 
F = {1\over 2}\sum_{i=1}^N \nabla\!s_i u_i^2
&&-\lambda_1 \left( \sum_{i=1}^N \nabla\!s_i u_i-{1\over{2}}\right) 
\nn\\
&&-\lambda_2 \left(
\sum_{i=1}^N \nabla\!s_i^2 u_i-{1\over{3}}\right ) 
\ea
gives 
\be
u_i=\ld_1{{\nabla\!s_i}\over{\nabla\!s_i}}+\ld_2 {{\nabla\!s_i^2}\over{\nabla\!s_i}}=\lambda_1+\lambda_2(s_i+s_{i-1}). 
\label{uform} 
\ee 
Imposing (\ref{cone}) and (\ref{ctwo}) determines $\lambda_1$ and
$\lambda_2$,
\ba
\lambda_1+\lambda_2&=&{1\over{2}},
\label{lone}\\
\lambda_1+\lambda_2+g\lambda_2&=&{1\over{3}},
\label{ltwo}
\ea
where $g$ defined by
\be
\sum_{i=1}^N {{\nabla\!s_i^2\nabla\!s_i^2}\over{\nabla\!s_i}}=1+g,
\label{gdef}
\ee
is given by
\be
g=\sum_{i=1}^N s_is_{i-1}(s_i-s_{i-1}).
\label{gsec}
\ee
By substituting in $s_is_{i-1}=(s_i^2+ s_{i-1}^2-(s_i-s_{i-1})^2)/2$,
one discovers that
$$
g=-\ft12 g+\ft12(1-\delta g),
$$
and therefore 
\be
g={1\over 3}(1-\delta g),\quad{\rm where}\quad
\delta g=\sum_{i=1}^N t_i^3.
\label{delg}
\ee
The factor ${1/3}$ is the continuum limit ($N\rightarrow\infty$) 
of $g$ when the sum is replaced by the integral $\int_0^1 s^2 ds$.
The evaluation of general sums of the form (\ref{gsec}) will be
further discuss below. This exact form for $g$ obviated the need 
to determine $g$'s upper bound as it is done originally in the work of 
Suzuki\cite{suzukinogo}, and in the more recent work on symplectic 
correctors\cite{chincor}.)
With $\lambda_1$ and $\lambda_2$ known, the minimium of $F$ is given by
\ba
F&=&{1\over 2}(\lambda_1+\lambda_2)^2+{1\over 2}g\lambda_2^2\nonumber\\
 &=&{1\over 4}+ {1\over {72g}}={1\over 6}
 +{1\over {24}}{{\delta g}\over{(1-\delta g})},
\label{min} 
\ea
and therefore,
\be
e_{VTV}\le -{1\over {24}}{{\delta g}\over{(1-\delta g})}.
\label{etvtf}
\ee
This implies that, first, 
$e_{VTV}$ must be negative. Secondly, its magnitude is
\be
|e_{VTV}|\ge {1\over {24}}{{\delta g}\over{(1-\delta g})}.
\label{lbound}
\ee
The Sheng-Suzuki theorem now follows as a 
simple corollary. If all the $t_i$'s are positive, then $e_{VTV}$ 
cannot vanish because its lower bound (\ref{lbound}), which depends 
on $\delta g$ as given by (\ref{delg}), cannot vanish. The only
way to achieve a fourth order forward algorithm is to keep
the commutator $[V,[T,V]]$ with coefficient $e_{VTV}$, but move it
to the left hand side of (\ref{prodform}). This means that for all such 
fourth order algorithms, the sum of factorization coefficients of 
all the $[V,[T,V]]$ terms must be positive. All such fourth order
algorithms are characterized by their respective values of $e_{VTV}$, 
and how well they saturate the lower bound (\ref{lbound}). Note that
in deriving this lower bound, we did not need to incorporate
the primary constraints $u_1=1$. 

A very different ``elementary" proof of the Sheng-Suzuki result has
been offered by Blanes and Casa\cite{blanes03}. Our work is
more precise in demonstrating that, not only $e_{VTV}$
cannot vanish, it has a lower bound (\ref{lbound}) determined
only by $\{t_i\}$.
   
Note also that $v_i=u_i-u_{i+1}$ and (\ref{uform}) implies that
\be
v_i=\lambda_2(s_{i-1}-s_{i+1})={1\over 2}{{(t_i+t_{i+1})}\over{(1-\delta g})}.
\label{vform}
\ee
Thus, if one insists that $e_{VTV}$ be zero, then $\delta g$ can
be zero only if at least one $t_i$ is negative such that
$(t_i+t_{i+1})$ or $(t_i+t_{i-1})$ remains negative. 
Eq.(\ref{vform}) then implies that its adjacent 
values of $v_i$ or $v_{i-1}$ must also be negative. Thus a fourth order
factorization without keeping any additional operator such as $[V,[T,V]]$
must have at least one pair of negative ${t_i,v_i}$ coefficients. 
This result was first proved by Goldman and 
Kaper\cite{goldman}. This simpler proof follows the idea 
of Blanes and Casa\cite{blanes03}.

\section{The sixth order case}

By incorporating the potential-like operator $[V,[T,V]]$, many
families\cite{chinchen02,ome02,ome03} of fourth order forward
algorithms have been found. They are not only indispensable
for solving time-irreversible
problems\cite{fchinl,fchinm,auer,ochin}; they are also superior to
existing fourth order algorithms in solving time-reversible
classical\cite{chin,chinchen03,ome02,ome03} and
quantum\cite{chinchen01,chinchen02} dynamical problems. It is
therefore of great interest to determine whether there exist
practical forward algorithms of even higher order. 
We show in this section that sixth order forward
algorithms requires the inclusion of the commutator $[V,[T,[T,[T,V]]]]$. 
The inclusion of $[V,[T,V]]$ which make possible fourth order forward
algorithms, is insufficient to guarantee a sixth order forward
algorithm. In general, if $F_{2n}(\ep)$ is a
$2n$th order forward decomposition of $\e^{\ep(T+V)}$, then
$F_{2n+2}(\ep)$ would require the inclusion of a new operator not
previously included in the construction of $F_{2n}(\ep)$. We have
proved the case of $n=1$ in the last section. The new operator
being 
\be 
V_1\equiv[V,[T,V]]. 
\label{vtv} 
\ee

Consider now the case $n=2$. In the following discussion, 
we will use the condensed bracket notation:
$[V^2T^3V]\equiv[V,[V,[T,[T,[T,V]]]]]$, etc.. We have shown in the
last section that, for positive ${t_i}$, with $u_i$ satisfying constraints
(\ref{cone}) and (\ref{ctwo}), we can factorize $\e^{\ep(T+V)}$
up to the form
\ba
\prod_{i=1}^N
\e^{t_i\ep T}\e^{v_i\ep V}
&=&\exp\biggl[\ep\Bigl( T+V+e_{VTV}\ep^2[VTV]\nn\\
&&\qquad+\ep^4\sum_{i=1}^4 e_iQ_i +O(\ep^6) \Bigr)\biggr],
\label{ffour}
\ea
where $e_{VTV}$ cannot be made to vanish, and $Q_i$ are
four independent operators described below. There is one error operator
$[TV]$ in first order, two error operators
$[TTV]$ and $[VTV]$ in second order, four operators $[TTTV]$, $[VTTV]$,
$[TVTV]$ and $[VVTV]$ in third order, and eight operators
\ba
&&[TTTTV],\quad [VTTTV],\quad [TVTTV],\quad [VVTTV],\nonumber\\
&&[TTVTV],\quad [VTVTV],\quad [TVVTV],\quad [VVVTV],
\nonumber
\ea
in fourth order. These error operators
are results of concatenating $T$ and $V$ with lower order
operators on the left. In each order, not all the operators are
independent. For example, setting $C=[AB]$
in the Jacobi identity
$$
[ABC]+[BCA]+[CAB]=0,
$$
gives $[ABC]=[BAC]$ and therefore
$$
[ABAB]=[BAAB].
$$
For the case where $[VTV]$ commutes with $V$ we also have
$[V^nVTV]=0$. Hence there are only two independent operators
$[TTTV]$, $[TVTV]$ in third order and four operators $[TTTTV]$,
$[VTTTV]$, $[TTVTV]$, $[VTVTV]$ in fourth order. The last two are
just $[TTV_1]$ and $[VTV_1]$, which resemble second order errors
for a new potential $V_1$. To have a sixth order algorithm, one
must eliminate these four error terms. Since $[TTV_1]$ and $[VTV_1]$ 
are linear in $V_1$, they can always be eliminated by including 
sufficient number of $V_1$ operators in the factorization process. 
The remaining error terms $[T^4V]$ and $[VT^3V]$ are unaffected by 
$V_1$ and can {\it only be eliminated by the choice of coefficients} 
$\{t_i,v_i\}$. Thus we can apply our previous strategy of dealing 
{\it only} with coefficients $\{t_i,v_i\}$ but now computing the 
error coefficient $e_{VT^3V}$ explicitly.

A careful reexamination of our proof for the Sheng-Suzuki theorem
shows that we have proved more than that's required. 
The minimization procedure produces a lower bound for $e_{VTV}$,
whereas the Sheng-Suzuki theorem only requires that 
$e_{VTV}$ not be zero. The expansion (\ref{uform}) merely served as 
a vehicle for demonstrating that, for any  $\{u_i\}$ 
satisfying (\ref{cone}) and (\ref{ctwo}), $e_{VTV}$
cannot vanish for positive $\{t_i\}$. We do not really need 
to minimize anything, or to determine an actual lower bound.  
This suggests a simple strategy for proving the sixth order case. 
It is sufficient to show that $e_{VT^3V}$ cannot vanish for any set 
of $\{u_i\}$ satisfying higher order constraints.  

\section{Proving the sixth order case}

As discussed in the last section, for a sixth order algorithm, 
a symmetric factorization must satisfy,
in additional to (\ref{cone}) and (\ref{ctwo}), the constraint 
(\ref{eodd}) for $n=2$,
\be
\sum_{i=1}^N\nabla\!s_i^3 u_i
={1\over{4}}.
\label{cthr}
\ee
Also, since the operator $T^4V$ uniquely tracks the commutator $[T^4V]$,
the error coefficient $e_{T^4V}$ will vanish if the expansion coefficient
of $T^4V$ is 1/5!. This means that factorization coefficients 
$\{t_i,v_i\}$ must also obey
\be
\sum_{i=1}^N\nabla\!s_i^4 u_i
={1\over{5}}.
\label{cfour}
\ee
These four constraints (\ref{cone}), (\ref{ctwo}),
(\ref{cthr}), and (\ref{cfour}), can be satisfied 
by the expansion,
\be
u_i=\ld_1+\ld_2 {{\nabla\! s^2_i}\over{\nabla\! s_i}}
         +\ld_3 {{\nabla\! s^3_i}\over{\nabla\! s_i}}
         +\ld_4 {{\nabla\! s^4_i}\over{\nabla\! s_i}}.
\label{usys}
\ee
We must now demonstrate that in this case,
$e_{VT^3V}$ cannot vanish if $\{t_i\}$ are all positive.

When $u_i$ is expanded via (\ref{usys}),
the four constraints (\ref{cone}), (\ref{ctwo}),
(\ref{cthr}), and (\ref{cfour}) produce the following 
set of four linear equations for $m=1$ to 4, 
\be
\sum_{n=1}^4 G_{mn}\ld_n={1\over{m+1}}.
\label{bgeq}
\ee
The matrix $G_{mn}$ is given by
\ba
G_{mn}=\sum_{i=1}^N {{\nabla\! s^m_i\nabla\! s^n_i}\over{\nabla\! s_i}}
&=&1+\sum_{i=1}^N s_is_{i-1}
{{\nabla\! s^{m-1}_i \nabla\! s^{n-1}_i}\over{\nabla\! s_i}}\nn\\
&\equiv& 1+g_{mn},
\label{gmatrix}
\ea
where we have used the identify
\be
{{\nabla\! s^{m}_i \nabla\! s^{n}_i}\over{\nabla\! s_i}}
=\nabla\! s^{m+n-1}_i+
s_is_{i-1}{{\nabla\! s^{m-1}_i \nabla\! s^{n-1}_i}\over{\nabla\! s_i}},
\label{ident}
\ee
to define the reduced symmetric matrix $g_{mn}$. Since $G_{1n}=G_{n1}=1$
(and hence $g_{1n}=g_{n1}=0$),
we can subtract the first constraint equation
\be
\ld_1+\ld_2+\ld_3+\ld_4={1\over 2}\label{coneone}
\ee
from the other three and reduce the system down to three equations
for $m=2$ to 4:
\be
\sum_{n=2}^4 g_{mn}\ld_n={1\over{m+1}}-{1\over 2}.
\label{geq}
\ee

By writing,
$s_i=s_{i-{1\over 2}}+{1\over 2}\nabla\! s_i$ and
$s_{i-1}=s_{i-{1\over 2}}-{1\over 2}\nabla\! s_i$ where
$s_{i-{1\over 2}}={1\over 2}(s_i+s_{i-1})$, we can 
systematically expand
\ba
&&\nabla\! s_i^n=(s_{i-{1\over 2}}+{1\over 2}\nabla s_i)^n
-(s_{i-{1\over 2}}-{1\over 2}\nabla s_i)^n\nonumber\\
 &&={{n!}\over{1!(n-1)!   }}s^{n-1}_{i-{1\over 2}}(\nabla\! s_i)
   +{{n!}\over{3!(n-3)!2^2}}s^{n-3}_{i-{1\over 2}}(\nabla\! s_i)^3
    +\cdots.
\nonumber
\ea
When each summant ${\nabla\! s^m_i\nabla\! s^n_i}/\nabla\! s_i$ is
expanded and compared with the similarly expanded integral 
$$
\int_{s_{i-1}}^{s_i} mns^{m+n-2}ds
={{mn}\over{m+n-1}}\nabla\!{s_i}^{m+n-1},
$$
we deduce that 
\ba
&&G_{mn}={{mn}\over{m+n-1}}\nn\\
&&-{1\over {12}}mn(m-1)(n-1)\Bigg\{
\sum_{i=1}^N s_{i-{1\over 2}}^{m+n-4}(\nabla\!s_i)^3\nonumber\\
&&\qquad\qquad\qquad+\,A_5
\sum_{i=1}^N s_{i-{1\over 2}}^{m+n-6}(\nabla\!s_i)^5+\cdots\Biggl\},
\label{gexpd}
\ea
with 
$$
A_5={1\over {120}}[(m+n-4)^2+(m-2)(2m-7)+(n-2)(2n-7)].
$$
The constant part of the matrix is the continuum limit 
($N\rightarrow\infty$) of the sum, which is the integral    
$$
\int_0^1 mns^{m+n-2}ds={{mn}\over{m+n-1}}.
$$
We will denote this constant part of the matrix as $G_{mn}^0$. 
The corresponding continuum
part of $g_{mn}$ is $g_{mn}^0=G_{mn}^0-1$.
The remaining finite parts of $G_{mn}$ in (\ref{gexpd}), which 
depends explicitly on $s_i$, will be denoted as $\delta G_{mn}$. 
Since $g_{mn}$ differs from $G_{mn}$ only by a constant, its finite part
$\delta g_{mn}$ is the same as that of $G_{mn}$, {\it i.e.},
$\delta g_{mn}=\delta G_{mn}$. By repeated applications of the identity
(\ref{ident}), one can reduce $g_{mn}$ to a sum of terms of the form
\be
\kappa(l,n)=\sum_{i=1}^N (s_is_{i-1})^l\nabla\! s^n_i.
\label{kap}
\ee 
Since the explicit form of $g_{mn}$ is known via (\ref{gexpd}), these
functions are not particularly useful as calculational tools. However,
they are very useful in quickly identifying the matrix element of 
$g_{mn}$ when doing analytical calculations. For later reference,
we list below some $g_{mn}$'s in terms of $\kappa(l,n)$: 
\ba
g_{22}&=&\kappa(1,1)\nonumber\\
g_{23}&=&\kappa(1,2)\nonumber\\
g_{24}&=&\kappa(1,3)\nonumber\\
g_{32}&=&\kappa(1,3)+\kappa(2,1)\label{kaplist}\\
g_{33}&=&\kappa(1,4)+\kappa(2,2)\nonumber\\
g_{34}&=&\kappa(1,5)+\kappa(2,3)+\kappa(1,3).
\nonumber
\ea  
Note that $g_{22}$ is the $g$ function of the last section.
From the general formula (\ref{gexpd}), one finds indeed that
$g_{22}^0=1/3$ and   
\be
\delta g_{22}=-{1\over 3}\sum_{i=1}^N(\nabla s_i)^3=-{1\over 3}\delta g.
\label{dgtwo}
\ee 

If we only keep the continuum matrix $g^0_{mn}$ in (\ref{geq}) 
$$
	  \pmatrix{
	  {1\over 3}&{1\over 2}&{3\over 5}\cr
	  {1\over 2}&{4\over 5}&{1}\cr		
	  {3\over 5}&{1}       &{9\over 7}\cr}
	  \pmatrix{\ld_2\cr\ld_3\cr\ld_4}
	 =\pmatrix{-{1\over 6}\cr -{1\over 4}\cr -{3\over 10}\cr}, 
$$  
the solution is trivial: $\ld_2=-{1\over 2}$, $\ld_3=0$, $\ld_4=0$. 
This suggests that we should also expand each $\ld_i$ into 
its continuum and finite part: $\ld_2=-{1\over 2}+\delta\!\ld_2$,
$\ld_3=\delta\!\ld_3$, $\ld_4=\delta\!\ld_4$. For our purpose, it is enough
to keep the leading finite size correction term, {\it i.e.}, we can neglect
the terms of the form $\delta g_{mn}$ $\delta\!\ld_{k}$. In this case, we just
have
\be
      \pmatrix{
      {1\over 3}&{1\over 2}&{3\over 5}\cr
      {1\over 2}&{4\over 5}&{1}\cr
      {3\over 5}&{1}       &{9\over 7}\cr}
      \pmatrix{\delta\!\ld_2\cr\delta\!\ld_3\cr\delta\!\ld_4}
     =\pmatrix{{1\over 2}\delta g_{22}\cr
               {1\over 2}\delta g_{23}\cr
               {1\over 2}\delta g_{24}\cr}.
\label{remat}
\ee
We do not need to solve each $\delta\!\ld_k$ explicitly; 
we only need to know that they are proportional to $\delta g_{2n}$. 
Since $\ld_1+\ld_2+\ld_3+\ld_4={1\over 2}$, this also implies that
$\ld_1=1+\delta\!\ld_1$ with
\be
\delta\!\ld_1+\delta\!\ld_2+\delta\!\ld_3+\delta\!\ld_4=0.
\label{deltaone}
\ee

The above discussion suggests that one should also separate $u_i$ into its 
continuum and finite part,
\be
u_i=(1-{1\over 2}{{\nabla\! s^2_i}\over{\nabla\! s_i}} )+\delta u_i.
\label{newu}
\ee
The constraints on $u_i$ now translate into constraints on $\delta u_i$:
\ba
\sum_{i=1}^N\nabla\!s_i^n\delta u_i&=&{1\over{n+1}}
-\sum_{i=1}^N\nabla s_i^n(1-{1\over 2}{{\nabla\! s^2_i}\over{\nabla\! s_i}} )
\nonumber\\
&=&{1\over{n+1}}-(1-{1\over 2}G_{2n})
={1\over 2}\delta g_{2n}
\label{delcont}
\ea
Recall that since $g_{1n}=g_{n1}=0$, we also have $\delta g_{n1}=\delta g_{1n}=0$.
The above constraints for $\delta u_i$ is exact. We have not yet invoked any particular
representation for $\delta u_i$. 

To illustrate how this formalism will be used, let's recompute the quadratic form
of the last section: 
\ba
&&\sum_{i=1}^N\nabla\! s_iu_i^2
=\sum_{i=1}^N\nabla\! s_i
\Bigl[(1-{1\over 2}{{\nabla\! s^2_i}\over{\nabla\! s_i}} )
    +\delta u_i\Bigr]^2\nonumber\\
	&=&{1\over 4}\sum_{i=1}^N{{\nabla\! s^2_i\nabla\! s^2_i}\over{\nabla\! s_i}}
	+2\sum_{i=1}^N\nabla\! s_i\,\delta u_i
	-\sum_{i=1}^N\nabla\! s^2_i\delta u_i +O(\delta u_i^2)\nn\\
	\label{delgs}\\
&=&{1\over 4}G_{22}-{1\over 2}\delta g_{22}
={1\over 3}-{1\over 4}\delta g_{22}={1\over 3} + {1\over 12}\delta g.
\label{chk} 
\ea
This then implies that 
\be
e_{VTV}= -{1\over {24}}\sum_{i=1}^Nt_i^3.
\label{evtvfp}
\ee
The first key observation is Eq.(\ref{delgs}): to leading order in 
$\delta g_{2n}$, this quadratic form only depends on the first 
two constraints on $\delta u_i$. Its leading finite part is unchanged
by additional, higher order constraints on $\delta u_i$. That is,
$\delta u_i$ can be very general. By inspection, $e_{VTV}$ above 
cannot vanish for positive $\{t_i\}$. Thus
this leading order calculation, while not sufficient to determine 
the exact lower bound for $e_{VTV}$, it is sufficient to show that
$e_{VTV}$ cannot vanish, and thus proving the 
Sheng-Suzuki theorem. 

Secondly, if $\delta u_i$ were to be
represented as
\be
\delta u_i=\delta\!\ld_2 ({{\nabla\! s^2_i}\over{\nabla\! s_i}}-1)
          +\delta\!\ld_3 ({{\nabla\! s^3_i}\over{\nabla\! s_i}}-1)
          +\delta\!\ld_4 ({{\nabla\! s^4_i}\over{\nabla\! s_i}}-1),
\label{delu}
\ee
then in order for the constraints (\ref{delcont}) to determine 
$\delta\!\ld_k$ to the same leading order in $\delta g_{2n}$ as in 
(\ref{remat}) it is enough to compute only the constant (continuum) 
part of any sums multiplying $\delta\!\ld_k$. This implies that we may replace any such sum by its 
integral, or by any other sum having the same integral. 
{\it Thus for any sum multiplying $\delta u_i$, we may replace it by 
another sum having the same integral}. This crucial simplification 
makes it unnecessary to solve for each $\ld_k$ explicitly.

To compute the error coefficient $e_{VT^3V}$, one must use an operator that
tracks the commutator $[VT^3V]$ uniquely. The analogous operator $T^3V^2$, 
whose expansion coefficient is easy to compute, is no longer suitable. 
Let $C_{T^3V^2}$ denote its expansion coefficient in terms of $\{t_i,v_i\}$
from the left-hand-side of (\ref{ffour}). By matching the same operator's
expansion coefficient from the right-hand-side, one finds\cite{forbert} 
\be
C_{T^3V^2}={1\over{5!}}-{1\over{3!}}e_{VTV}-e_{T^2VTV}-e_{VT^3V}.
\label{nogood}
\ee
It is difficult to disentangle $e_{VT^3V}$ from the contaminating effects of
$e_{VTV}$ and $e_{T^2VTV}$. The three operators that track $[VT^3V]$ uniquely
are $VT^3V$, $VT^2VT$, and $TVT^2V$. We choose the symmetric choice $VT^3V$,
whose coefficient is related to $e_{VT^3V}$ by
\be
C_{VT^3V}={1\over{5!}}+2e_{VT^3V}.  
\label{good}
\ee
From the left hand side of (\ref{ffour}), one deduces
\be
C_{VT^3V}={1\over{3!}}\sum_{i=1}^{N-1}v_i\sum_{j=i+1}^N(s_j-s_i)^3v_j.
\label{cnod}
\ee
This quadratic form in $\{v_i\}$ is difficult to work with because it is not
diagonal in $u_i$ or some other variables. In the Appendix, we show that it
can be simplified to the following bi-diagonal form,
\be
C_{VT^3V}={1\over{3!}}\left(
3\sum_{i=1}^N\nabla\! s_i z_i^2-\sum_{i=1}^N\nabla\! s_i^3u_i^2-{1\over 4}\right),
\label{bidia}
\ee
where $z_i$ is defined by
\be
z_i=\sum_{j=i}^N v_j s_j.
\label{zvar}
\ee
The required coefficient $e_{VT^3V}$ can now be computed from
\be
e_{VT^3V}={1\over{12}}\left(
3\sum_{i=1}^N\nabla\! s_i z_i^2-\sum_{i=1}^N\nabla\! s_i^3u_i^2-{3\over {10}}\right).
\label{evtttv}
\ee  

The quadratic form involving $u_i^2$ is
\ba
&&\sum_{i=1}^N\nabla\! s_i^3u_i^2
=\sum_{i=1}^N\nabla\! s_i^3(1-{1\over 2}{{\nabla\! s^2_i}\over{\nabla\! s_i}})^2
+2\sum_{i=1}^N\nabla\! s_i^3\delta u_i\nn\\
&&\qquad\qquad\qquad\qquad
-{3\over 2}\sum_{i=1}^N\nabla\! s^4_i\delta u_i +O(\delta u_i^2)\label{eqsum}\\
&&\qquad={3\over 4}-G_{32}+{1\over 4}(G_{33}+G_{24})+\delta g_{23}-{3\over 4}\delta g_{24}
\nonumber\\
&&\qquad={1\over{10}}+{1\over 4}\delta g_{33}-{1\over 2}\delta g_{24}.
\label{usquare}
\ea
In (\ref{eqsum}), we have replaced the sum involving
$\nabla\! s^3_i\nabla\! s^2_i/\nabla\! s_i$ by its integral equivalent
$(3/2)\nabla\! s^4_i$. Also, we have used
the identity
$${{\nabla\! s^3_i}\over{\nabla\! s_i}}
\left({{\nabla\! s^2_i\nabla\! s^2_i}\over{\nabla\! s_i}}\right)
={{\nabla\! s^3_i\nabla\! s^3_i}\over{\nabla\! s_i}}
+{{\nabla\! s^4_i\nabla\! s^2_i}\over{\nabla\! s_i}}-\nabla\! s_i^5
$$

Given the expansion (\ref{usys}) for $u_i$, we can deduce 
the corresponding expansion for $z_i$. From (\ref{zvar}),
we can rewrite $z_i$ as
\be
z_i=u_is_i+\sum_{j=i+1}^Nu_j\nabla s_j.
\label{zrewr}
\ee
For $u_i=\lambda_n\nabla s_i^n/\nabla s_i$, we have
\ba
z_i&=&\lambda_n\Bigl[\ft{\nabla s_i^n}{\nabla s_i} s_i+(1-s_i^n)
\Bigr],\nn\\
&=&	\lambda_n\Bigl[(s_i^{n-1}+s_{i-1}\ft{\nabla s_i^{n-1}}{\nabla s_i}) s_i+(1-s_i^n)
\Bigr],\nn\\
&=&	\lambda_n\Bigl[1+s_is_{i-1}\ft{\nabla s_i^{n-1}}{\nabla s_i}
\Bigr].
\ea
Hence corresponding to (\ref{usys}), 
$z_i$ has the expansion
\ba
z_i&=&\ld_1+\ld_2(1+s_is_{i-1})
         +\ld_3(1+s_is_{i-1}{{\nabla\! s^2_i}\over{\nabla\! s_i}})\nn\\
 &&\qquad\qquad\qquad+\ld_4(1+s_is_{i-1}{{\nabla\! s^3_i}\over{\nabla\! s_i}}).
\label{zsys}
\ea
One can check that this form for $z_i$ satisfies the four constraints
(\ref{cone}), (\ref{ctwo}), (\ref{cthr}), and (\ref{cfour}) when they 
are expressed in terms of $z_i$:
$$
z_1=\ld_1+ \ld_2+ \ld_3+ \ld_4={1\over 2},
$$
and for $m=1$ to 3,
\be 
\sum_{i=1}^N \nabla\! s_i^m z_i={1\over{m+2} }.
\label{zcons}
\ee
The identity (\ref{ident}) is needed to show that (\ref{zcons})	
is equivalent to the last three constraint equations for $u_i$. 
As in the case of $u_i$, we can write $z_i$ in the form
\be
z_i={1\over 2}(1-s_is_{i-1})+ \delta z_i
\label{newz}
\ee
and transfer the last three constraints on $z_i$ to $\delta z_i$,
\be
\sum_{i=1}^N\nabla\! s_i^{n-1}\delta z_i={1\over 2}\delta g_{2n}.
\label{constz}
\ee
The quadratic form for $z_i$ is then
\ba
&&\sum_{i=1}^N\nabla\! s_iz_i^2
	={1\over 4}\sum_{i=1}^N\nabla\!s_i(1-s_is_{i-1})^2
	+\sum_{i=1}^N\nabla\! s_i\delta z_i\nn\\ 
&&\qquad\qquad-\sum_{i=1}^N\nabla\! s_i(s_is_{i-1})\delta z_i 
	 +O(\delta z_i^2)
	\nonumber\\
&=&{1\over 4}-{1\over 2}\kappa(1,1)+{1\over 4}\kappa(2,1)
+{1\over 2}\delta g_{22}
-{1\over 3}\sum_{i=1}^N\nabla\! s^3_i\delta z_i
	\nonumber\\
&=&{1\over 4}-{1\over 2}g_{22}+{1\over 4}(g_{33}-g_{24})
+{1\over 2}\delta g_{22}
-{1\over 6}\delta g_{24}  \nonumber\\
&=&{2\over {15}}+{1\over 4}\delta g_{33}
-{5\over {12}}\delta g_{24}
\label{evalz} 
\ea
We have again replaced the sum involving $\nabla\!s_i(s_is_{i-1})$
by its integral equivalent $(1/3)\nabla\!s_i^3$ and used 
(\ref{kaplist}) to express the required sum in terms of $g_{mn}$'s.
Thus the bi-diagonal form is
$$
3\sum_{i=1}^N\nabla\! s_i z_i^2-\sum_{i=1}^N\nabla\! s_i^3u_i^2
=
{3\over {10}}+{1\over 4}(2\delta g_{33}
-3\delta g_{24}).
$$
From (\ref{gexpd}) we find,
\ba
\delta g_{33}=-3\sum_{i=1}^Ns_{i-{1\over 2}}^2(\nabla\!s_i)^3
-{1\over 20}\sum_{i=1}^N(\nabla\!s_i)^5\nonumber\\
\delta g_{24}=-2\sum_{i=1}^Ns_{i-{1\over 2}}^2(\nabla\!s_i)^3
-{1\over 10}\sum_{i=1}^N(\nabla\!s_i)^5,
\label{twog}
\ea
and therefore finally,
\be
e_{VT^3V}={1\over{240}}\sum_{i=1}^N(\nabla\!s_i)^5
={1\over{240}}\sum_{i=1}^Nt_i^5\,\,!
\label{efinal}
\ee
This is remarkably similar to (\ref{evtvfp}). 
Thus if $\{t_i\}$ are all positive, then $e_{VT^3V}$ cannot vanish.
No sixth order positive factorization scheme is possible without
including the commutator $V_3=[VT^3V]$.

\section{Beyond Sixth Order}

In Sections II, we have shown that in order to have
a fourth order forward algorithm, one must include the commutator
$V_1=[VTV]$ in the factorization process. In the last section, we 
have proved that in order to have a sixth order forward algorithm 
one must include in addition to $V_1$, the commutator $V_3=[VT^3V]$.
By repeating the same argument, it is not difficult to
discern the pattern of higher order forward algorithms. In going
from the (2n)$^{\rm th}$ to the (2n+2)$^{\rm th}$ order, one must add a new commutator 
$$
V_{2n-1}=[VT^{2n-1}V]
$$
to the factorization process.
A proof of this general result is a straightforward generalization
of our approach in the last section, but technically much more involved.  
For example, to prove the eighth order case,
we must track $e_{VT^5V}$ uniquely via 
the operator $VT^5V$'s coefficient given by $S_5/5!$, 
where $S_5$ as shown in the Appendix, is tri-diagonal in $u_i$, $z_i$ and
$$
y_i=\sum_{j=i}^{N}v_js^2_j.
$$
One then has to work out the expansion for $y_i$ as in the case of
$z_i$. Moreover, since $e_{VT^5V}$ is anticipated to be 
$\propto \sum_{i=1}^N(\nabla\!s_i)^7$, one can no longer
ignore contribution of order $(\delta u_i)^2\propto 
(\sum_{i=1}^N(\nabla\!s_i)^3)^2$. Thus the current formalism, while  
powerful in determining $e_{VTV}$ variationally and $e_{VT^3V}$ 
perturbatively, is too demanding for the general case. To prove
such a general result, one must find a less explicit approach.  

\section{Conclusions}

In this work, we have introduced a framework for analyzing and
understanding the structure of factorized algorithms. 
There are three key ideas: 1) The order constraints and error 
coefficients can be tracked by operators and expressed directly in 
terms of factorization coefficients. 2) By introducing a suitable 
representation for the factorization coefficients, the order 
constraints and error terms can be solved analytically. 3) For many 
purposes, it  is sufficient to determine the error coefficients 
perturbatively. This last point is specially important. All previous 
works on factorization algorithms are based on exact decompositions. 
Since this is difficult to do analytically, one can make little 
progress except numerically . This work shows that 
a leading order calculation is sufficient to establish most of the 
important results we know about these algorithms. In particular, 
we have provided a constructive proof of the Shang-Suzuki theorem.
Most importantly, we have shown that in order to have a sixth order 
forward time step algorithm, one must include the commutator 
$[VT^3V]$ in the factorization process. 
   
This work suggests that there is regularity to the existence
of forward algorithms. In order to have only positive time steps,
one must continue to enlarge one's collection of constituent
operators for factorizing $\e^{\ep(T+V)}$. For a (2n)$^{\rm th}$ order
forward algorithm one must include all commutators of the form
$[VT^{2k-1}V]$ from $k=1$ to $k=n-1$, in addition to $T$ and $V$.
The proof of this general result is currently beyond scope of our
perturbative approach. Moreover, the massive cancellations that 
produced the sixth order result (\ref{efinal}) strongly suggest 
that a better formulation, with these cancellations built-in, 
must be possible. This work suggests that a more powerful way of 
understanding the structure of these algorithms is still waiting 
to be found.
 
The need to include $[VT^3V]$ make it difficult to construct, 
but does not necessarily preclude the possibility of a sixth order 
forward algorithm. One simply has to work harder to devise practical
ways of obtaining $[VT^3V]$ without computing it directly. 
Work is currently in progress
toward this goal.

\begin{acknowledgments}

I thank Harald Forbert for pointing out the inadequacy of an 
earlier version of this work and for many stimulating discussions. 
This work is supported, in part, by a National Science Foundation 
grant, No. DMS-0310580.

\end{acknowledgments}

\appendix
\section{Coefficient of $VTV$, $VT^3V$ and $VT^5V$}

There is a systematic way of diagonalizing the sum 
$$
S_m=\sum_{i=1}^{N-1} \sum_{j=i+1}^{N}v_i(s_j-s_i)^mv_j 
$$
needed in computing the error coefficients $e_{VT^mV}$.
The above is a sum over the upper triangle of a $N\times N$ 
square matrix and can be denoted more simply 
as $\sum_{j>i}$.

The general form we need to diagonalize is
\be
S(f,g)=\sum_{j>i}f_i(g_j-g_i)f_j
=\sum_{i>j}f_j g_i f_i-\sum_{j>i}f_ig_if_j
\label{sfg}    
\ee 
where we have interchanged the summation indices in the first term
on the right-hand-side. 
The key point here is that if we introduce a new variable
$$
h_i=\sum_{j=i}^{N}f_j,
$$
such that $f_i=h_i-h_{i+1}$, then the second term on the right 
hand side of (\ref{sfg}) is only a single sum. The first term 
can be eliminated by completing the``square matrix". Let
$\sum_{i}f_ig_i=P$ and $\sum_jf_j=F$ be known sums, then
\be
P F=\sum_if_ig_i\sum_j f_j=\sum_i f^2_i g_i
+\sum_{i>j}f_ig_i f_j+\sum_{j>i}f_ig_i f_j.
\label{fgsq}
\ee
Subtracting (\ref{sfg}) from (\ref{fgsq}) gives
\ba
P F&-&S(f,q) = \sum_i f^2_i g_i+2\sum_{j>i}f_ig_i f_j\nonumber\\
&=&\sum_{i=1}^Ng_i(h_i-h_{i+1})^2+2\sum_{i=1}^Ng_i(h_i-h_{i+1})h_{i+1}\nonumber\\
&=&\sum_{i=1}^Ng_i(h_i^2-h_{i+1}^2)=\sum_{i=1}^N\nabla\! g_ih^2_i,
\label{diag}
\ea
and hence,
\be
S(f,g)=P F-\sum_{i=1}^N\nabla\! g_ih^2_i
\label{fsfg}
\ee
For the case of $m=1$, we have $f_i=v_i$, $g_i=s_i$, $h_i=u_i$, $F=1$ from (\ref{tvcon}), 
and $P=({1\over2}+e_{TV})$ from (\ref{etv}). Therefore, we have
$$
S_1=({1\over2}+e_{TV})-\sum_{i=1}^N \nabla\!s_i u_i^2.
$$
Since the coefficient of $VTV$ is just $S_1={1\over 3!}+e_{VTV}$, the
above is identical to (\ref{evtv}). The use of the more complicated
operator $VTV$ determines the same $e_{VTV}$, as it must. 
 
For $m=3$, we have
$$
S_3=\sum_{j>i}v_i(s_j^3-s_i^3)v_j-3\sum_{j>i}v_is_i(s_j-s_i)s_jv_j 
$$
Assuming now that all linear constraints on $v_i$ are satisfied up to
the relevant order, we have for the first and second term on the right
respectively, 
$f_i=v_i$, $g_i=s^3_i$, $h_i=u_i$, 
$F=1$, $P={1\over 4}$ and $f_i=s_iv_i$, $g_i=s_i$, 
$h_i=z_i$, $F={1\over 2}$, a$P={1\over 3}$. Hence we have
$$
S_3={1\over 4}-\sum_{i=1}^N\nabla\! s^3_iu^2_i
-3\Bigl( {1\over 6}-\sum_{i=1}^N\nabla\! s_iz^2_i\Bigr),
$$
where
$$
z_i=\sum_{j=i}^{N}v_js_j.
$$
The coefficient of $VT^3V$ is $S_3/3!$. Since $[VT^3V]$ contains the operator
$VT^3V$ twice, we have
$$
{1\over 6}S_3={1\over {5!}}+2\, e_{VT^3V},
$$
and therefore
\be
12\,e_{VT^3V}=S_3-{1\over {20}}=3\sum_{i=1}^N\nabla\! s_iz^2_i
-\sum_{i=1}^N\nabla\! s^3_iu^2_i-{3\over 10}
\label{vt3v}
\ee

For the case $m=5$, we have  
\ba
&&S_5=\sum_{j>i}v_i(s_j^5-s_i^5)v_j
-5\sum_{j>i}v_is_i(s^3_j-s^3_i)s_jv_j \nn\\
&&\qquad\qquad\qquad\quad+10\sum_{j>i}v_is^2_i(s_j-s_i)s^2_jv_j 
\ea
For the first term we have
$f_i=v_i$, $g_i=s^5_i$, $h_i=u_i$, 
$F=1$, and $P={1\over 6}$. For the second term we have $f_i=s_iv_i$, $g_i=s^3_i$, 
$h_i=z_i$, $F={1\over 2}$, and $P={1\over 5}$. For the third term, we have
$f_i=s^2_iv_i$, $g_i=s_i$, 
$h_i=y_i$, $F={1\over 3}$, and $P={1\over 4}$. We therefore have
\ba
S_5&=&{1\over 6}-\sum_{i=1}^N\nabla\! s^5_iu^2_i
-5\Bigl( {1\over {10}}-\sum_{i=1}^N\nabla\! s_i^3z^2_i\Bigr)\nn\\
&&\qquad\qquad\qquad+10\Bigl( {1\over{12}}-\sum_{i=1}^N\nabla\! s_i y^2_i\Bigr)\nonumber\\
&=&{1\over 2}-\sum_{i=1}^N\nabla\! s^5_iu^2_i
+5\sum_{i=1}^N\nabla\! s_i^3z^2_i
-10\sum_{i=1}^N\nabla\! s_i y^2_i,\nn
\ea
where
$$
y_i=\sum_{j=i}^{N}v_js^2_j.
$$

\newpage
\centerline{REFERENCES}


\begin{thebibliography}{10}
\bibitem{yoshi} H. Yoshida, Celest. Mech. {\bf 56},27 (1993).
\bibitem{hairer}{\it Geometric Numerical Integration}, by E. Hairer,
                C. Lubich, and G. Wanner, 
				Springer-Verlag, Berlin-New York, 2002.
\bibitem{mcl02} R. I. McLachlan and G. R. W. Quispel, Acta Numerica,
               {\bf 11}, 241 (2002).
\bibitem{chinchen03}S. A. Chin, and C. R. Chen,
                  ``Forward Symplectic Integrators for Solving Gravitational
                    Few-Body Problems", arXiv, astro-ph/0304223.
\bibitem{hirono}T. Hirono, W. Lui, S. Seki, Y. and Yoshikuni, 
                IEEE Trans. Mirco. Theory and Tech., {\bf 49}, 1640 (2001).
\bibitem{ti}M. Takahashi and M. Imada,
          J. Phys. Soc. Jpn {\bf 53}, 3765 (1984).
\bibitem{chincor}S. A. Chin
                  ``Quantum Statistical Calculations and Symplectic
                  Corrector Algorithms", arXiv, cond-mat/0312021.
\bibitem{feit}D. Feit, J. A. Fleck, Jr., and A. Steiger, 
          J. Comput. Phys. {\bf 47}, 412 ( 1982)
\bibitem{chinchen01}S. A. Chin and C. R. Chen,
                J. Chem. Phys. {\bf 114}, 7338 (2001).
\bibitem{chinchen02}S. A. Chin and C. R. Chin,
                   J. Chem. Phys. {\bf 117}, 1409 (2002).
\bibitem{sheng}Q. Sheng, IMA J. Num. Anaysis, {\bf 9},
              199 (1989).
\bibitem{suzukinogo}M. Suzuki, J. Math. Phys. {\bf 32}, 400 (1991).
\bibitem{suz95}M. Suzuki, Phys. Lett. {\bf A 201}, 425 (1995).
\bibitem{suzfour}M. Suzuki, {\it Computer Simulation Studies in
            Condensed Matter Physics VIII},
           eds, D. Landau, K. Mon and H. Shuttler (Springler, Berlin, 1996).
\bibitem{chin} S.A. Chin, Physics Letters {\bf A} 226, 344 (1997).
\bibitem{ruth83}R. Ruth, IEEE Transactions
           on Nuclear Science, {bf 30}, 2669 (1983).
\bibitem{fchinl}H. A. Forbert and S. A. Chin,
                Phys. Rev. {\bf E 63}, 016703 (2001).
\bibitem{fchinm}H. A. Forbert and S. A. Chin,
                Phys. Rev. {\bf B 63}, 144518 (2001).
\bibitem{auer}J. Auer, E. Krotscheck, and S. A. Chin,
                J. Chem. Phys. {\bf 115}, 6841 (2001).
\bibitem{ochin}O. Ciftja and S. A. Chin, Phys. Rev.
               {\bf B 68}, 134510 (2003).
\bibitem{jang} S. Jang, S. Jang and G. A. Voth,
              J. Chem. Phys. {\bf 115} 7832, (2001).
\bibitem{ome02}I. P. Omelyan, I. M. Mryglod and R. Folk,
               Phys. Rev. {\bf E66}, 026701 (2002).
\bibitem{ome03}I. P. Omelyan, I. M. Mryglod and R. Folk,
               Comput. Phys. Commun. {\bf 151} 272 (2003)
\bibitem{blanes03}S. Blanes and F. Casas,`` On the existence of positive 
                coefficients for operator splitting schemes of order higher 
				than two", preprint GIPS 2003-004, http://www.focm.net/gi/gips
\bibitem{goldman}D. Goldman and T. J. Kaper, SIAM J. Numer. Anal.,{ \bf 33}, 
                   349 (1996).
\bibitem{forbert}Harald Forbert, private communications.

\end{thebibliography}
\end{document}